\documentclass[aps,prl,twocolumn,superscriptaddress]{revtex4-1}

\renewcommand{\thefootnote}{\fnsymbol{footnote}}
\renewcommand{\theequation}{\arabic{section}.\arabic{equation}}

\usepackage{amssymb}
\usepackage{amsfonts}
\usepackage{amsmath}
\usepackage{bm}
\usepackage{textcomp}
\usepackage{color}
\usepackage{longtable}
\usepackage{graphicx}
\usepackage{dcolumn}

\usepackage{epsfig}

\begin{document}
\renewcommand{\thefootnote}{\fnsymbol{footnote}}
\renewcommand{\theequation}{\arabic{section}.\arabic{equation}}

\title{Origin of low-temperature magnetic ordering in Ga$_{1-x}$Mn$_x$N}

\author{M. Sawicki} \email{mikes@ifpan.edu.pl}
\affiliation{Institute of Physics, Polish Academy of Sciences,
  al.~Lotnik\'ow 32/46, 02-668 Warszawa, Poland}

\author{T. Devillers} \email{thibaut.devillers@jku.at}
\affiliation{Institut f\"ur Halbleiter- und
  Festk\"orperphysik, Johannes Kepler University, Altenbergerstr.~69,
 4040 Linz, Austria}

\author{S. Ga{\l}{\c{e}}ski}
\affiliation{Institute of Physics, Technical University of {\L}\'od\'z, ul.~W\'olcza\'nska 215, 90-924 {\L}\'od\'z, Poland}

\author{C. Simserides} \affiliation{Physics Department, University of Athens,
Panepistimiopolis, Zografos, 15784 Athens, Greece}

\author{S. Dobkowska}
\affiliation{Institute of Physics, Polish Academy of Sciences,
  al.~Lotnik\'ow 32/46, 02-668 Warszawa, Poland}

\author{B. Faina} \affiliation{Institut f\"ur Halbleiter- und
  Festk\"orperphysik, Johannes Kepler University, Altenbergerstr.~69,
  4040 Linz, Austria}

\author{A. Grois} \affiliation{Institut f\"ur Halbleiter- und
  Festk\"orperphysik, Johannes Kepler University, Altenbergerstr.~69,
4040 Linz, Austria}

\author{A. Navarro-Quezada} \affiliation{Institut f\"ur
  Halbleiter- und Festk\"orperphysik, Johannes Kepler University,
  Altenbergerstr.~69, 4040 Linz, Austria}

\author{K. N. Trohidou}
\affiliation{Institute of Materials Science, NCSR Demokritos, 15310 Athens, Greece}

\author{J. A. Majewski}
\affiliation{Institute of Theoretical Physics, Faculty of Physics, University of Warsaw, 00-681 Warszawa, Poland}

\author{T. Dietl}
\affiliation{Institute of Physics, Polish Academy of Sciences,
  al.~Lotnik\'ow 32/46, 02-668 Warszawa, Poland}
\affiliation{Institute of Theoretical Physics, Faculty of Physics, University of Warsaw, 00-681 Warszawa, Poland}

\author{A. Bonanni} \email{alberta.bonanni@jku.at}
\affiliation{Institut f\"ur Halbleiter- und Festk\"orperphysik,
  Johannes Kepler University, Altenbergerstr.~69, 4040 Linz,
  Austria}

\begin{abstract}
By employing highly sensitive millikelvin SQUID magnetometry, the magnitude of the Curie temperature as a function of the Mn concentration $x$ is determined for thoroughly characterized Ga$_{1-x}$Mn$_x$N. The interpretation of the results in the frame of tight binding theory and of Monte Carlo simulations, allows us to assign the spin interaction to ferromagnetic superexchange and to benchmark the accuracy of state-of-the-art {\em ab initio} methods in predicting the magnetic characteristics of dilute magnetic insulators.

\end{abstract}

\date{\today}

\maketitle

The extensive studies of dilute magnetic semiconductors (DMSs) and oxides over the last decade \cite{Ohno:2010_NM,Pearton:2004_JPC} have persistently confronted the researchers with experimental and conceptual challenges, making the field to be one of the most controversial in today's condensed matter physics. More specifically, it has become increasingly clear that the premise of dilute magnetic alloys -- where the magnetic constituents incorporate randomly and substitutionally into the host crystal -- breaks entirely down in a number of systems \cite{Bonanni:2010_CSR}. In particular, the distribution of magnetic ions is often non-uniform and the ions tend to occupy also interstitial positions. Moreover, the paramount importance of disorder, defects, and strong correlation makes that the theoretical and computational modeling of these materials has often been misleading. However, recently some consensus on the {\em ab initio} tools appropriate to study the (ferro)magnetism in these systems has been reached \cite{Sato:2010_RMP,Zunger:2010_P}.

Since GaN and its alloys with Al and In have already realized their potential in photonics and high power electronics, reaching the status of the technologically most significant semiconductor materials next to Si, the addition of magnetism opens wide application prospects. In particular, the presence of ferromagnetic interactions {\em without} band carriers, together with a sizable spin splitting of the excitonic states revealed already for Ga$_{1-x}$Mn$_x$N \cite{Bonanni:2011_PRB,Suffczynski:2011_PRB}, indicates the suitability of this system for magnetooptical devices, such as optical isolators, circumventing the destructive effect of antiferromagnetic interactions specific to II-VI Mn-based DMSs \cite{Zayets:2005_JOSA_B}. Surprisingly, however, in previous works a variety of different magnetic behaviors is reported for Ga$_{1-x}$Mn$_x$N at the same nominal Mn concentration $x$. This compound was found by some groups to be non-magnetic \cite{Zajac:2001_APL}, whereas according to others it shows either low-temperature spin-glass freezing \cite{Dhar:2003_PRB} or ferromagnetism with a Curie temperature $T_{\text{C}}$ ranging from 8\,K \cite{Sarigiannidou:2006_PRB} up to over 300\,K \cite{Pearton:2004_JPC,Bedair:2010_IEEE_S}.

In this Letter we report on studies of magnetic hysteresis down to millikelvin temperatures for Ga$_{1-x}$Mn$_x$N epitaxial layers, in which the high crystallinity, the random distribution of Mn ions, and the extremely weak degree of compensation by residual donors were assessed by a range of electron microscopy, synchrotron radiation, optical, and magnetic resonance techniques \cite{Bonanni:2011_PRB}.  The magnetic phase diagram $T_{\text{C}}(x)$  established in this way allows us to verify the predictive power of {\em ab initio} methods.  With the support of tight-binding theory and Monte Carlo simulations, we corroborate the experimental results and find that state-of-the-art first principles approaches overestimate the magnitude of the Mn--Mn exchange energies by an order of magnitude. In this way, we signify that dilute magnetic insulators constitute a relevant system to benchmark newly developed tools for computational design of functional magnetic materials.

\begin{figure*}[th]
  \centering
  \centerline{\includegraphics[width=5cm]{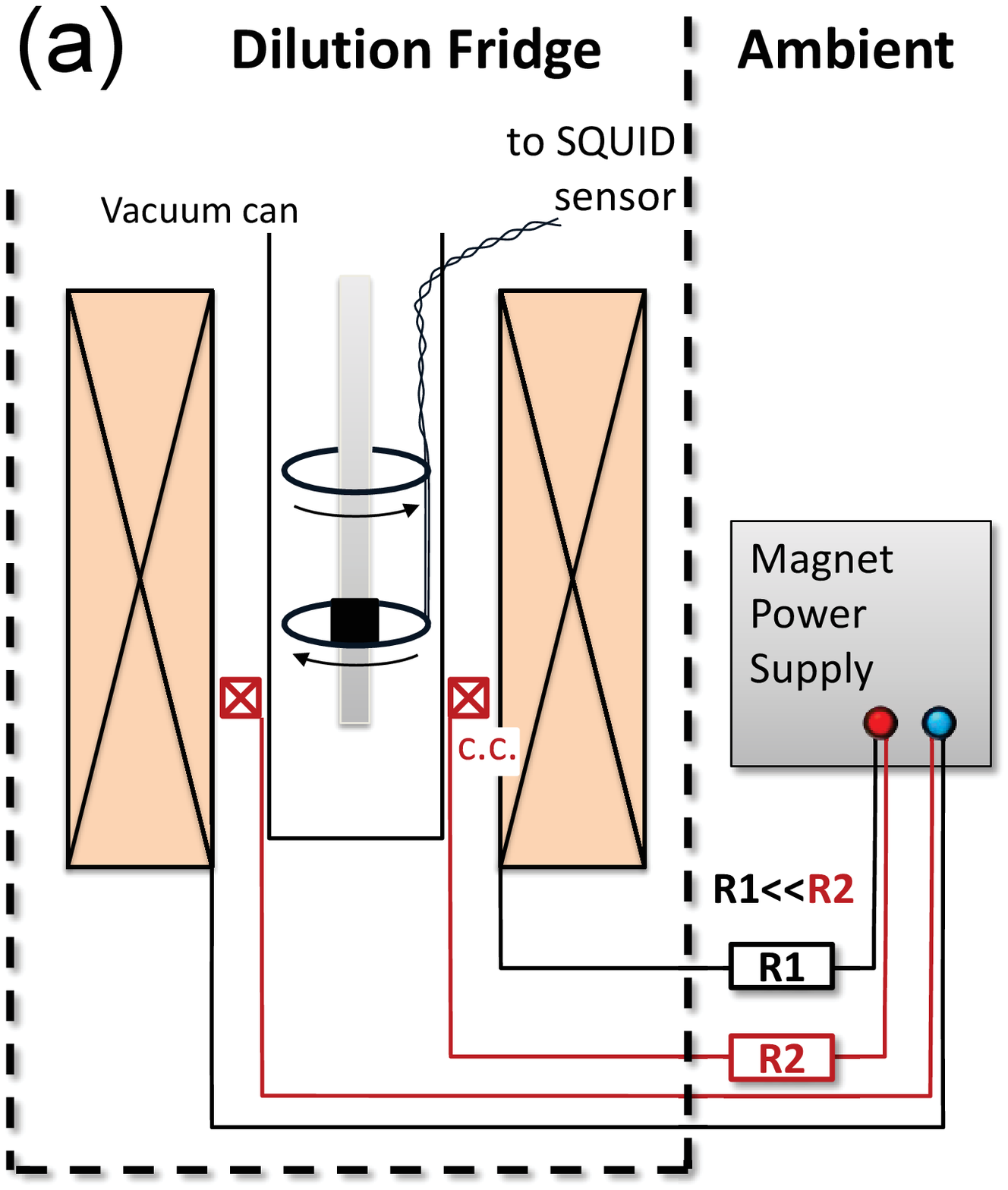},\includegraphics[width=12.5cm]{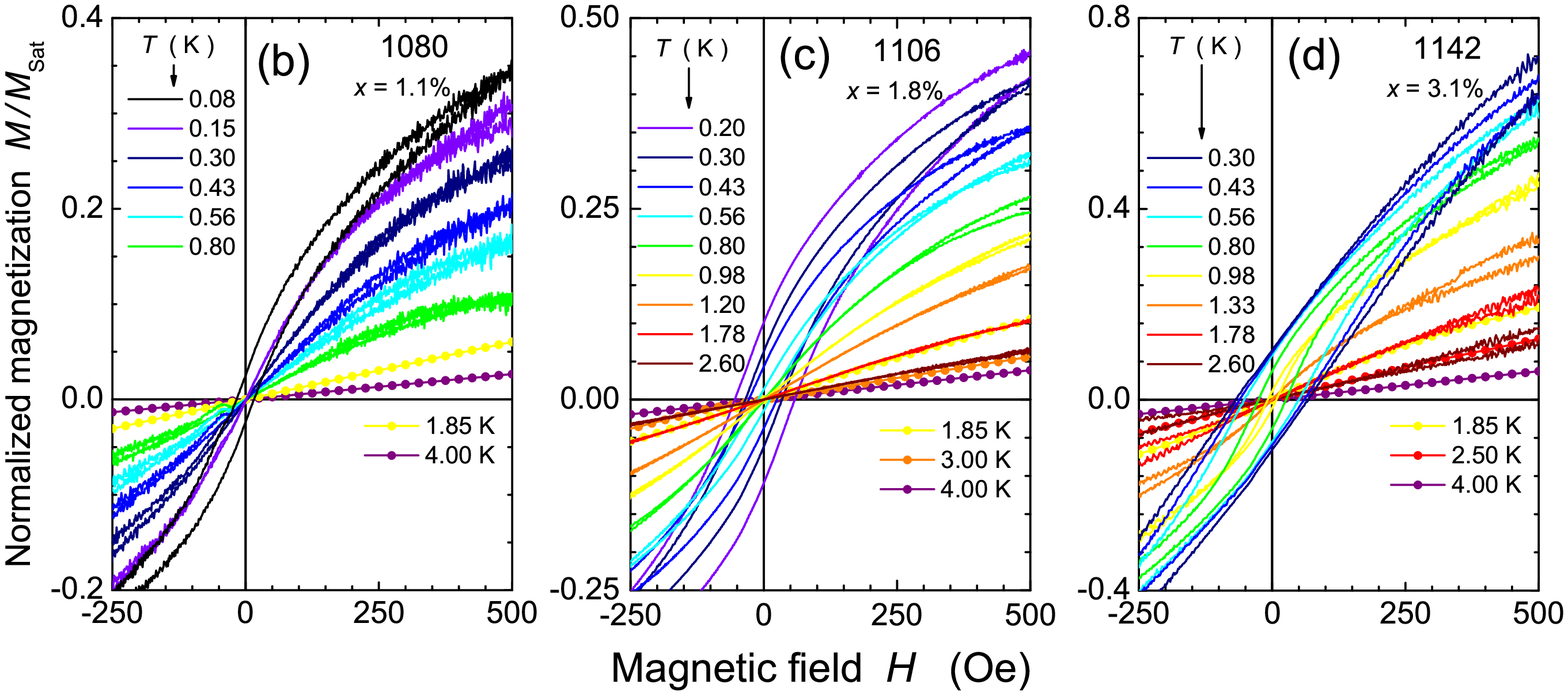}}
  \caption{(Color online) (a) Sketch of the mK magnetometer employed for this work. For maximum linearity of the set-up the superconducting pick-up coils are made of pure niobium wire and the magnetic field is generated by a copper wound electromagnet immersed in a helium dewar. An additional copper-made flux-balancing coil (c.~c.) is placed asymmetrically with respect to the main coil and is connected to the same power supply through a resistive current divider (R1 and R2) adjusted to obtain an adequately flat magnetic field (reference) response when a reference undoped GaN layer is placed inside the pick-up. The whole set-up has been calibrated against the known magnetic moment of a signal coil located at the sample position.  (b)-(d) Magnetic hysteresis loops at various temperatures for Ga$_{1-x}$Mn$_x$N measured in the dilution fridge set-up (lines) and in a commercial SQUID magnetometer (bullets), respectively. The reported values of the magnetization are normalized to their magnitudes at 70\,kOe and at 1.85\,K.}
  \label{fig:mK_data}
\end{figure*}

The samples discussed here have been grown by metalorganic vapor phase epitaxy according to the procedure reported previously \cite{Stefanowicz:2010_PRBa,Suffczynski:2011_PRB,Bonanni:2011_PRB}. In particular, Ga$_{1-x}$Mn$_x$N has been deposited
onto GaN/c-sapphire at a substrate temperature of 850$^{\circ}$C. In order to maximize the
homogeneous and substitutional incorporation of Mn, to avoid phase separation, and at the same time to vary from sample to sample the actual concentration of Mn in a controlled way, the flow rate of the Ga precursor (TMGa) has been changed over the samples series from 5 to 1 standard cubic centimeters per minute (sccm), the temperature of the Mn precursor source (MeCp$_{2}$Mn) from 17$^{\circ}$C to 22$^{\circ}$C, while its flow rate has been maintained constant at 490\,sccm for all the samples considered.

The films have been thoroughly characterized by secondary-ion mass spectroscopy; high resolution
(scanning) transmission electron microscopy with capabilities allowing for chemical analysis, including energy-dispersive x-ray spectroscopy, high angle annular dark-field mode, and electron energy loss spectroscopy; high-resolution and synchrotron x-ray diffraction; synchrotron extended x-ray absorption fine-structure; synchrotron x-ray absorption near-edge structure; infrared optics; electron spin resonance; and superconducting quantum interference device (SQUID) \cite{Bonanni:2011_PRB}. This extensive analysis has allowed us to rule out the presence of Mn-precipitation and indicates that: (i) up to a Mn content of 3.1\% at least 95\% of the Mn ions has the charge state 3+ and is substitutionally incorporated in the host crystal; (ii) the layers are highly resistive even at 300~K indicating no charge transport \emph{via} band or Mn gap states. The concentration $x$ of Mn has been tuned from 0.5\% of cations (2$\times 10^{20}$\,cm$^{-3}$) to 3.1\% for different samples. Here, the layers with respectively $x$\,=\,1.1\% (sample 1080), 1.8\% (sample 1106) and 3.1\% (sample 1142) are considered, where the code of the sample numbers is the same adopted previously \cite{Bonanni:2011_PRB}.

Previous magnetization studies of these Ga$_{1-x}$Mn$_x$N films were carried out down to 1.8~K \cite{Bonanni:2011_PRB}, employing a commercial SQUID magnetometer and a measurement technique specifically developed in order to examine meaningfully thin layers of magnetically dilute semiconductors \cite{Sawicki:2011_SST}. The data pointed to the presence of ferromagnetic interactions between Mn spins but no signatures of long range magnetic order were found \cite{Bonanni:2011_PRB}. In order to extend the range of our measurements, we have installed in our $^3$He/$^4$He dilution fridge a homemade SQUID-based magnetometry set-up that allows for high-sensitivity detection of magnetic moments during continuous field sweeps up to about 800\,Oe and down to 20\,mK, as sketched in Fig.\,\ref{fig:mK_data}(a).

\begin{figure}[tb]
  \centering
  \includegraphics[width=8.cm]{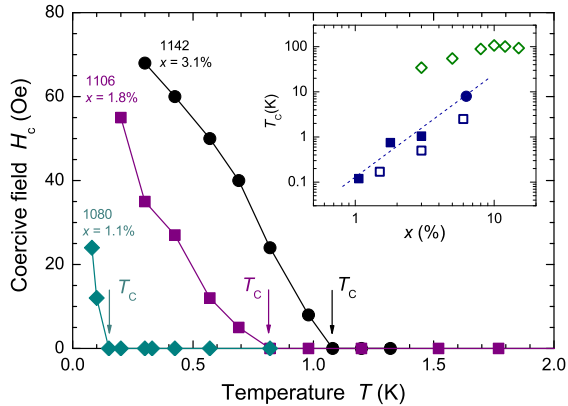}
  \caption{(Color online)  Magnitude of the coercive field as a function of temperature for the considered Ga$_{1-x}$Mn$_x$N samples (full points), employed to determine the Curie temperatures. Inset: experimental Curie temperatures as a function of Mn content $x$, together with the experimental result of Ref.~\onlinecite{Sarigiannidou:2006_PRB} (solid squares and circle, respectively). The dotted line shows the scaling dependence  $T_{\text{c}} \propto x^{m}$ with $m=2.2$.  The results of {\em ab initio} (Ref.~\onlinecite{Sato:2010_RMP}, open diamonds) and tight-binding (this work, open squares) approaches are also shown.}
  \label{fig:TC_determination}
\end{figure}

In Fig.~\ref{fig:mK_data}(b)-(d) the magnetization loops obtained at various temperatures for the studied films are reported. The emergence of a long-range ferromagnetic order is witnessed by the appearance of hystereses, whose width (and height) exhibit a critical behavior on lowering temperature, allowing to determine the magnitude of $T_{\text{C}}$ for particular films, as shown in  Fig.\,\ref{fig:TC_determination}. In contrast to the case of those dilute magnetic semiconductors and oxides which show high values of $T_{\text{C}}$ independent of the magnetic ion concentration $x$ \cite{Pearton:2004_JPC}, our data point to a significant variation of $T_{\text{C}}$ with $x$, as expected from any model of ferromagnetism within a system of randomly distributed localized spins \cite{Sato:2010_RMP,Dietl:2000_S}.

To put these data into a broader context, we note that over the last decade  Ga$_{1-x}$Mn$_x$N has reached the status of a model system, whose magnetic properties have been theoretically examined by over twenty different groups employing a variety of {\em ab initio} methods \cite{Sato:2010_RMP,Zunger:2010_P,Sandratskii:2004_PRB,Gonzalez:2011_PRB}. Notably, these approaches predict consistently the presence of ferromagnetic coupling, as observed. Moreover, again in agreement with experimental observations, a cross-over to antiferromagnetic interactions is expected upon donor compensation that leads to the reduction Mn$^{3+}$ $\rightarrow$ Mn$^{2+}$  \cite{Sandratskii:2004_PRB}. However, a quantitative comparison of the experimental and theoretical $T_{\text{C}}$ values as shown in Fig.~\ref{fig:TC_determination} implies that the state-of-the-art computational approaches overestimate the measured values by an order of magnitude.

To unravel this issue, we recall that uncompensated Ga$_{1-x}$Mn$_x$N can be classified as a dilute magnetic insulator \cite{Bonanni:2011_PRB}, in which the absence of electrons and holes makes carrier-mediated spin-spin coupling \cite{Zener:1951_PRa} irrelevant, and the lack of mixed valence -- all magnetic ions are in the same 3+ charge state -- precludes the presence of double exchange \cite{Zener:1951_PRb}. In this situation, $superexchange$ accounts for spin-spin interactions \cite{Anderson:1950_PR}. Its sign is determined by the Anderson-Goodenough-Kanamori rules, whereas the character of the $p-d$ hybridization controls the magnitude of the scaling exponent $m$ describing the dependence of the magnetic critical temperature on $x$, $T_{\text{C}}(x) \propto x^m$.  The fact that the scaling law with similar exponents is obeyed by spin glass freezing in II-VI DMSs ($m=1.9 \pm 0.1$) \cite{Twardowski:1987_PRB} and by ferromagnetic ordering in Ga$_{1-x}$Mn$_x$N ($m =2.2 \pm 0.3$) -- as evidenced in Fig.~\ref{fig:TC_determination} -- strongly supports the superexchange scenario.

In order to evaluate the sign and magnitude of the spin-spin coupling, we adopt for Ga$_{1-x}$Mn$_x$N an experimentally constrained procedure developed by one of us and co-workers for II-VI compounds doped with transition metals (TM) \cite{Blinowski:1996_PRB}. Within this approach, the magnetic ions are described in terms of the Parmenter's \cite{Parmenter:1973_PRB} generalization of the Anderson hamiltonian for the relevant electronic configuration of the TM taking into account the Jahn-Teller distortion \cite{Gosk:2005_PRB,Stroppa:2009_PRB}, whereas the host band structure is modeled by the $sp^3s^*$ tight binding approximation, employing the established parametrization for GaN \cite{Ferhat:1996_pssb} in the cubic approximation. Other parameters of the model \cite{Blinowski:1996_PRB} are taken from experimental studies of optical \cite{Graf:2003_pssb} as well as photoemission and soft x-ray absorption spectroscopy \cite{Hwang:2005_PRB} of Ga$_{1-x}$Mn$_x$N. In particular, the charge transfer energy between the Mn ion and the top of the valence band, Mn$^{2+}  \rightarrow$ Mn$^{3+}$ is $e_1 = -1.8$\,eV  \cite{Graf:2003_pssb}, which together with the on site correlation energy for Mn$^{3+}$ ions  \cite{Graf:2003_pssb}, $U = 1$\,eV and the on site exchange energy for  Mn$^{2+}$ ions, $\Delta = E(S = 5/2) - E(S = 3/2) = 2$\,eV leads to $e_2 = 4.8$\,eV, where the uncertainty on the relevant energies, $e_1$ and $e_2$, is presumably of the order of $\pm$\,0.5\,eV. The magnitude of the $p$-$d$ hybridization energy is $V_{pd\sigma} = -1.5\pm 0.1$\,eV \cite{Hwang:2005_PRB}.

\begin{figure}[tb]
  \centering
  \includegraphics[width=8cm]{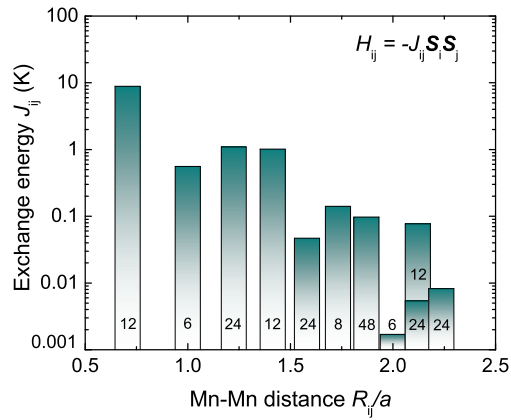}
  \caption{(Color online) Exchange energies provided by the tight binding model for Mn pairs in zinc blende GaN as a function of the distance between Mn spins in units of lattice parameter $a$. The numbers of equivalent cation sites at the particular distances $R_{\text{ij}}$ are also given.}
  \label{fig:Jn}
\end{figure}

Since the effect of spin-orbit splitting is small in the valence band of GaN, the spin dependent interaction between two Ga-substitutional Mn spins is described by a scalar Heisenberg coupling $H^{\gamma\delta}_{\text{ij}} = -J^{\gamma\delta}_{\text{ij}}\bm{S}_{\text{i}}\bm{S}_{\text{j}}$, where $\gamma$ and $\delta$ denote the one $t_{2g}$ orbital (either $xy, xz$ or $yz$) which is empty at the Mn$^{3+}$ ion i and j, respectively.  The magnitudes of $J^{\gamma\delta}_{\text{ij}}$ are evaluated within the fourth order perturbation theory in $V_{pd}$ for all possible orbital configurations $\gamma$ and $\delta$. Similarly to the case of Cr$^{2+}$ ions in II-VI compounds \cite{Blinowski:1996_PRB}, the main contribution originates from quantum hopping involving occupied $t_{2g}$ orbitals at the one Mn$^{3+}$ ion and the empty orbital at the other Mn$^{3+}$ ion. For the orbital configurations in question we find that the interaction is {\em ferromagnetic} at all distances.

In order to compare quantitatively the theoretical and experimental results, we assume a statistical distribution of directions corresponding to tetragonal Jahn-Teller distortions and determine an average value of the exchange energy $J_{\text{ij}}$ characterizing the coupling of Mn$^{3+}$ pairs at a given distance $R_{\text{ij}}$ in the fcc cation sublattice.  In this way we obtain the values of $J_{\text{ij}}$ shown in Fig.~\ref{fig:Jn}. By taking into account the coupling up to ten subsequent neighbor positions we allow for the formation of a percolation cluster down to $x\,\approx$\,1.2\% \cite{Osorio:2007_PRB}. Importantly, when compared to the {\em ab initio} results \cite{Sato:2010_RMP,Sandratskii:2004_PRB,Gonzalez:2011_PRB}, our values of $J_{\text{ij}}$ are significantly smaller and may, therefore, lead to a better agreement with the experimental data. This is substantiated by the $T_{\text{C}}$ values summarized in the inset to Fig.\,\ref{fig:TC_determination}, which have been obtained by employing Monte Carlo simulations and the cumulant crossing method \cite{Binder:1988_B}. We note here, that the magnitudes of $J_{\text{ij}}$ are rather sensitive to the input parameters. For instance, by changing $e_2$ from 4.8~eV to 4.4~eV, {\em i.\,e.}, within its expected uncertainty,  the computed $T_{\text{C}}$ values are in agreement with the experimental data.

In summary, the substantial agreement between our experimental values of $T_{\text{C}}(x)$ and our tight binding and Monte Carlo simulations, has made it possible to identify $\textit{ferromagnetic superexchange}$ as the microscopic mechanism accounting for the ferromagnetic interaction between localized spins in Ga$_{1-x}$Mn$_x$N. Because of its short range character, this coupling leads to rather low $T_{\text{C}}(x)$ values. Furthermore, the results allow to shed light on the predictive power of the current first principles methods, broadly employed to treat the case of Ga$_{1-x}$Mn$_x$N. We note, that the computed magnitudes of $T_{\text{C}}(x)$,  obtained from the available implementations of DFT, are much higher than the experimental values. This disagreement originates presumably from an overestimation --  inherent to the local spin density approximation -- of the metallization of the $d$ level, this being an error $not$ affecting the approach employed here. It would be now worth to assess, whether other {\em ab initio} methods, like $\textit{e.\,g.}$, hybrid density functionals \cite{Stroppa:2009_PRB} or a combination of DFT with a dynamic mean field approximation \cite{Jakobi:2011_PRB}, could bring computational results even closer to the experimental values.~\\

This work was supported by the FunDMS Advanced Grant of the ERC (Grant No. 227690) within the Ideas 7th Framework Programme of European Community, by the InTechFun (Grant No.
POIG.01.03.01-00-159/08), by the SemiSpinNet (Grant No. PITNGA-2008-215368), and by the Austrian FWF (P20065, P22477, P20550). Computer time in Athens was partly provided by the National Grid Infrastructure HellasGrid.

\end{document}